%% file: main_arxiv.tex
\documentclass[11pt]{article}

\usepackage[margin=0.7in]{geometry} 
\usepackage{setspace}
\usepackage[superscript]{cite} 
\usepackage{indentfirst} 
\usepackage[table]{xcolor}
\usepackage[pdftex]{graphicx}
\usepackage{hyperref}
\usepackage{tikz}
\usepackage{booktabs}
\usepackage{xltabular}
\usepackage{longtable}
\usepackage{xcolor}

\usepackage{comment}
\usepackage{tabularx}
\usepackage{pdflscape}
\usepackage{multirow}
\usepackage{caption}
\usepackage{longtable}
\usepackage{authblk}
\input{packages}

\title{\textbf{Impact of Large Language Model Assistance on Patients Reading Clinical Notes: A Mixed-Methods Study}}

\author[1,*]{Niklas Mannhardt}
\author[1,2,*,+]{Elizabeth Bondi-Kelly}
\author[3]{Barbara Lam}
\author[4]{Chloe O’Connell}
\author[1,6]{Hussein Mozannar}
\author[1]{Mercy Asiedu}
\author[1]{Alejandro Buendia}
\author[1]{Tatiana Urman}
\author[5]{Irbaz B. Riaz}
\author[1]{Catherine E. Ricciardi}
\author[1]{Monica Agrawal}
\author[1]{Marzyeh Ghassemi}
\author[1]{David Sontag}

\affil[1]{Massachusetts Institute of Technology}
\affil[2]{University of Michigan, Ann Arbor, MI}
\affil[3]{Division of Clinical Informatics, Division of Hematology \& Oncology, Department of Medicine, Beth Israel Deaconess Medical Center}
\affil[4]{Massachusetts General Hospital}
\affil[5]{Division of Hematology/Oncology, Department of Medicine, Mayo Clinic}
\affil[6]{Microsoft Research}

\affil[*]{Equal contribution}
\affil[+]{Corresponding author, \href{mailto:ecbk@umich.edu}{ecbk@umich.edu}}

\date{}

\begin{document}

\maketitle
\vspace*{-1cm}

\begin{abstract}
Large language models (LLMs) have immense potential to make information more accessible, particularly in medicine, where complex medical jargon can hinder patient comprehension of clinical notes. We developed a patient-facing tool using LLMs to make clinical notes more readable by simplifying, extracting information from, and adding context to the notes. We piloted the tool with clinical notes donated by patients with a history of breast cancer and synthetic notes from a clinician. Participants (N=200, healthy, female-identifying patients) were randomly assigned three clinical notes in our tool with varying levels of augmentations and answered quantitative and qualitative questions evaluating their understanding of follow-up actions. Augmentations significantly increased their quantitative understanding scores. In-depth interviews were conducted with participants (N=7, patients with a history of breast cancer), revealing both positive sentiments about the augmentations and concerns about AI. We also performed a qualitative clinician-driven analysis of the model's error modes.
\end{abstract}

\maketitle

\input{body}

\bibliographystyle{abbrv}
\bibliography{ref}

\clearpage

\appendix
\input{appendix}

\end{document}

%% file: packages.tex
\usepackage[table]{xcolor}
\usepackage{graphicx}
\usepackage{hyperref}

\usepackage{tikz}
\usepackage{booktabs}
\usepackage{xltabular}
\usepackage{longtable}

\usepackage{xcolor}
\usepackage{comment}
\usepackage{tabularx}
\usepackage{pdflscape}
\usepackage{multirow}
\usepackage{longtable}

\usepackage{enumitem}

%% file: body.tex
\input{files/introduction}

\input{files/methods}

\input{files/results}

\input{files/discussion}

%% file: files/introduction.tex
\section{Introduction}\label{ch:introduction}
\begin{figure}
    \centering
    \includegraphics[width=0.8\textwidth]{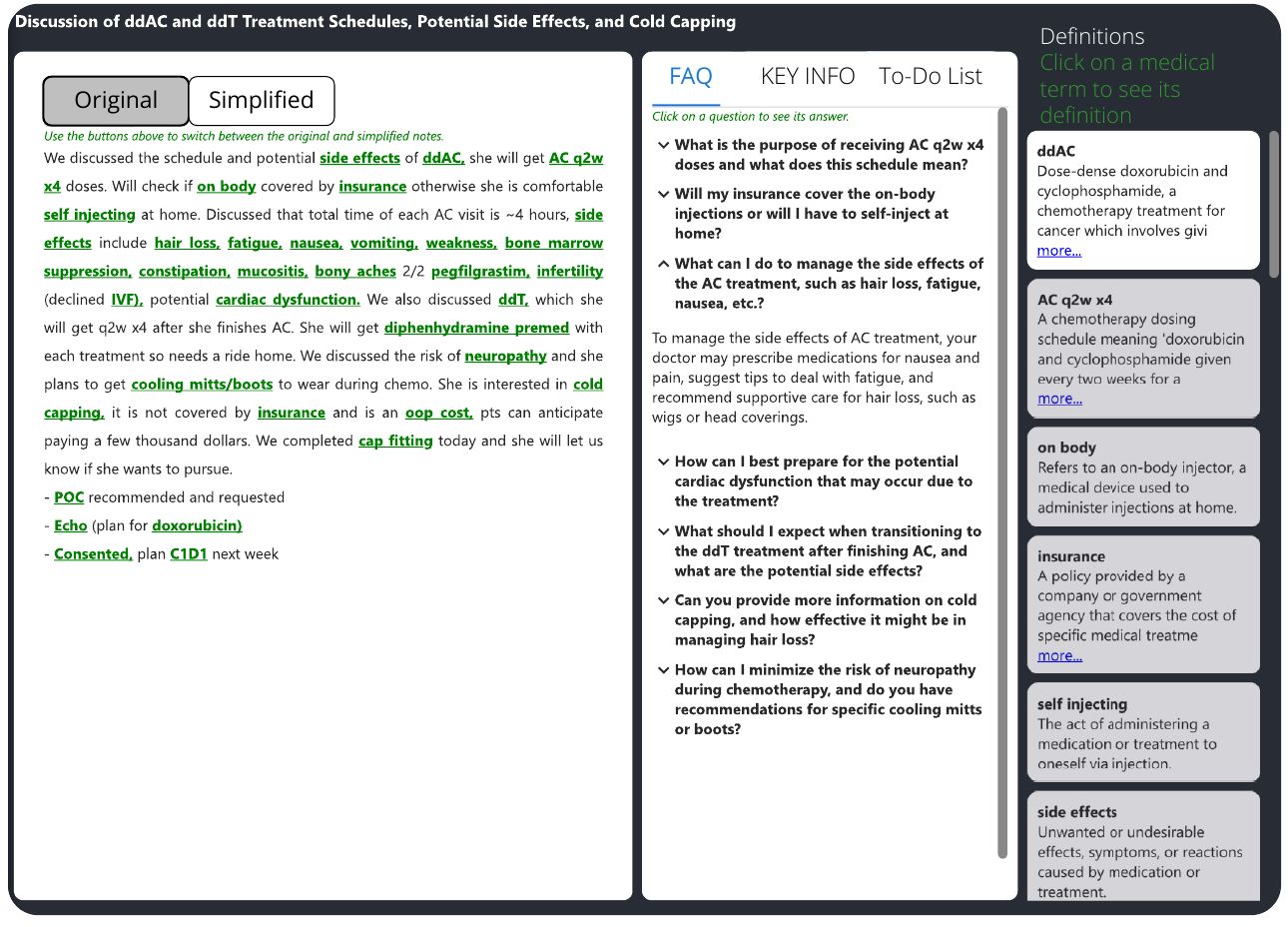}
    \caption{Our user interface to assist in the comprehension of clinical notes, containing (from left to right) an option to toggle between the original clinical note and simplified augmentation, an optional pane with the FAQ, key information, and to-do list augmentations, and finally, an optional definitions augmentation pane. Headings have been shown in the same formatting as visualizations in the rest of the paper.}
    \label{fig:intro_ui}
\end{figure}

The rapid advancement and improved performance of natural language processing could reduce barriers in communication and make information accessible to wider populations. Language models have the potential to help personalize and tailor the level and format of messaging to a variety of needs, e.g., to facilitate new avenues for public interest communication. Medicine, and public health more broadly, could particularly stand to benefit from new innovations in this space to facilitate improved patient-provider interaction \cite{bitterman2024promise}. 

The information needs of patients are not well addressed by existing clinical workflows. In appointments, clinicians frequently use unexplained medical terminology \cite{MILLER20221261}, which patients have trouble understanding \cite{10.1001/jamanetworkopen.2022.42972, SCHNITZLER2017112}. This information barrier has large implications; over a third of patients do not remember basic information about new medications \cite{tarn2011new}. 
In response, patients' rights advocates have pushed for increased transparency by enabling patients to read and access their own clinical notes, which act as physician-authored summaries of an encounter.  Reviewing clinical notes after an appointment can enable patients to better understand their medical conditions and improve care plan adherence \cite{shaverdian2019impact}. In 2021, the $\textrm{21}^{st}$ Century Cares Act mandated that providers make clinical notes digitally available to all patients \cite{blease2020opennotes, salmi2020clinicianviews}. 

Unfortunately, the clinical notes in electronic health records are written for the audiences of the clinicians themselves, other clinicians involved in care, and administrative, billing, and insurance stakeholders \cite{davidson2004s}. As patients are not the target audience, clinical notes are often filled with medical jargon that can be prohibitively difficult to understand. A recent study found it to be the most frequently reported barrier to reading clinical notes \cite{kambhamettu2024explainable}. Another study found that improving the coherence of notes improved patient recall, but the intervention was too time consuming to be reasonably scalable \cite{smith2011beyond}.
 
In this work, we studied whether large language models (LLMs) can help bridge this gap and make clinical notes more accessible. We focused on breast cancer, where comprehension can be particularly challenging, and the corresponding benefits of improved health literacy are significant \cite{shaverdian2019impact,alpert2019opennotes,salmi2020clinicianviews}. In particular, patients with cancer note a mismatch between the desired and actual levels of involvement in care. The core questions we sought to investigate were (1) whether and how AI augmentations enhance patients' experience with their clinical notes, and (2) whether there are negative impacts from these augmentations. Our contributions are as follows:
\begin{itemize}
    \item An end-to-end tool for improving clinical note comprehension, with five different LLM-enabled augmentations (see Figure \ref{fig:intro_ui}) 
    \item A quantitative survey study (N=200) on patient comprehension in which participants were exposed to three different levels of augmentations 
    \item In-depth interviews and qualitative analysis of seven patients with a history of breast cancer on their opinions and concerns about the interface and LLM augmentations
    \item An analysis of the types of errors induced by the LLM augmentations, as rated by three clinical partners
\end{itemize}

\section{Related Work}\label{ch:related}
Given the potential impact of increased accessibility to clinical notes, there have been several studies investigating the benefits of the status quo, current concerns, and possible mitigations. Several works have explored the existing impact of accessible clinical notes through interviews with patients and their care partners; benefits have included improved memory of the care plan, a better understanding of the medical condition and potential side effects, reassurance, and the ability for care teams to support patients remotely \cite{SHAVERDIAN2019102, kayastha, Esch, JACKSON2022290, desroches}. Further, the populations likelier to report benefits from reviewing their clinical notes are patients who are less educated, nonwhite, older, or with English as their nonprimary language \cite{walker2019seven_years}. However, many patients have made recommendations on how to improve the writing of clinical notes, citing difficulties in note comprehension due to the jargon in notes and not knowing where to focus attention while reading \cite{kambhamettu2024explainable, Leveille2020evaluate, kujala2022experiences}.

Given these known flaws in the current state of clinical notes, there have been suggestions for clinicians on how to improve note readability \cite{klein2016your,lam2023patient}. After access to notes, one study found that oncologists were writing slightly longer notes with increased readability, as measured by automatic scores \cite{rahimian}. Others found that manual interventions, such as adding definitions, improving coherence of text, and expanding acronyms, did improve patient understanding, as measured by patient questionnaires \cite{smith2011beyond, grossman-liu, khasawneh2022effect}. Unfortunately, the interventions were manual and difficult to scale, and the burden of documentation is already considered a leading cause of physician burnout \cite{10.1001/jama.2018.12777}. 

In response, the field of clinical natural language processing has pursued how to automate the simplification of clinical and biomedical text \cite{ondov2022survey}. Until a few years ago, methods have largely consisted of (i) rule-based methods, which are brittle and difficult to generalize, and (ii) deep learning methods that have suffered from operating in a low-resource setting with a lack of health data. These had seen limited success in clinical text \cite{ondov2022survey}. More recently, researchers have started exploring large language models as another avenue in preliminary studies for text simplification, in medicine and in the general domain \cite{kew2023bless}; other avenues have also been explored for patient communication including chatbots and question answering \cite{cai2023, ayers_physician_ai}. In the general domain, text simplification has often focused on shorter documents, e.g. sentence-level \cite{kew2023bless, wu2024depth}. Clinically, LLMs have been found to improve the readability of radiology reports as measured by automated metrics, though these have been found to only loosely correlate with patient comprehension \cite{li2023decoding, zheng2017readability}. Several works have found positive benefits from LLM text simplification in smaller user studies, as measured by patient perception \cite{makhmutova2024automated,schmidt2024simplifying}. In contrast, we conducted a much larger, patient-centric study, that complements qualitative perception with quantitative measures of patient comprehension of realistic medical action items for the patient.

Much of the benchmarking of LLMs in medicine has focused on multiple choice question answering. Meanwhile, while difficult to qualify, large language models still have flaws when conducting open-ended generation in medicine. Several papers have found the introduction of unsupported facts, sometimes referred to as \textit{hallucinations}, as a nontrivial problem in medical summarization \cite{cai_generation_2022}. In one study of open-ended medical question answering,  physicians rated responses across axes such as bias and medical knowledge \cite{singhal2023towards}. In our study, we studied a much more constrained scenario than open-ended question answering, as there is long context (the clinical note) that the response should be conditioned on, and we described a taxonomy of LLM errors for this longer-form clinical setting, based on thematic analysis.

%% file: files/methods.tex
\section{Clinical Notes Augmentations}\label{sec:methods}
\subsection{Augmentation Types}\label{subsec:augmentations}
We propose five types of augmentations to clinical notes to support patients. Each of the augmentations was obtained by passing a clinical note with a natural language instruction tailored for each augmentation as an input prompt to the LLM GPT-4 (specifically version GPT-4-0314) \cite{openai2023gpt4}. We employed prompt engineering strategies specific to each augmentation, primarily focusing on zero-shot prompting and prompt chaining \cite{wu}.


\begin{itemize}
    \item \textbf{Definitions}: recognizing and defining medical terms in clinical notes. The prompt was to  ``Make a list of the medical words that appear in this clinical note. Write each word as it appears in the note without modification. Define each of these terms. Definitions should be 1-2 sentences in simple English."
    \item \textbf{Simplification}: translating the original clinical note into layperson language. The prompt was to ``Rewrite this clinical note in simple English. Do not remove names of medications or diagnoses."
    \item \textbf{FAQ}: generating common questions and answers based on the clinical note. The prompt was to ``Make a list of 7 common questions a patient might ask their doctor after reading this clinical note. Answer the questions in 2-3 sentences using simple English."
    \item \textbf{Key Information}: extracting key information from the clinical note.  The prompt was to ``Make a list of the medications mentioned in this clinical note. Make a list of the upcoming appointments explicitly mentioned in this clinical note. If the note does not mention appointments, say ‘No Appointments’"
    \item \textbf{To-do List}: extracting any action items that the patient needs to remember. The prompt was to ``Make a todo list of tasks the patient is responsible for before their next appointment based on the clinical note. Tasks need to be specific and actionable. Tasks should not be recurring events. If there are no such tasks mentioned in the note, say 'No Tasks'"
\end{itemize}

Once we generated each of these augmentations, we integrated them into a user interface for patients to interact with to improve their experience reading clinical notes.

\subsection{User Interface}

\begin{figure}
    \centering
    \includegraphics[width=0.9\textwidth]{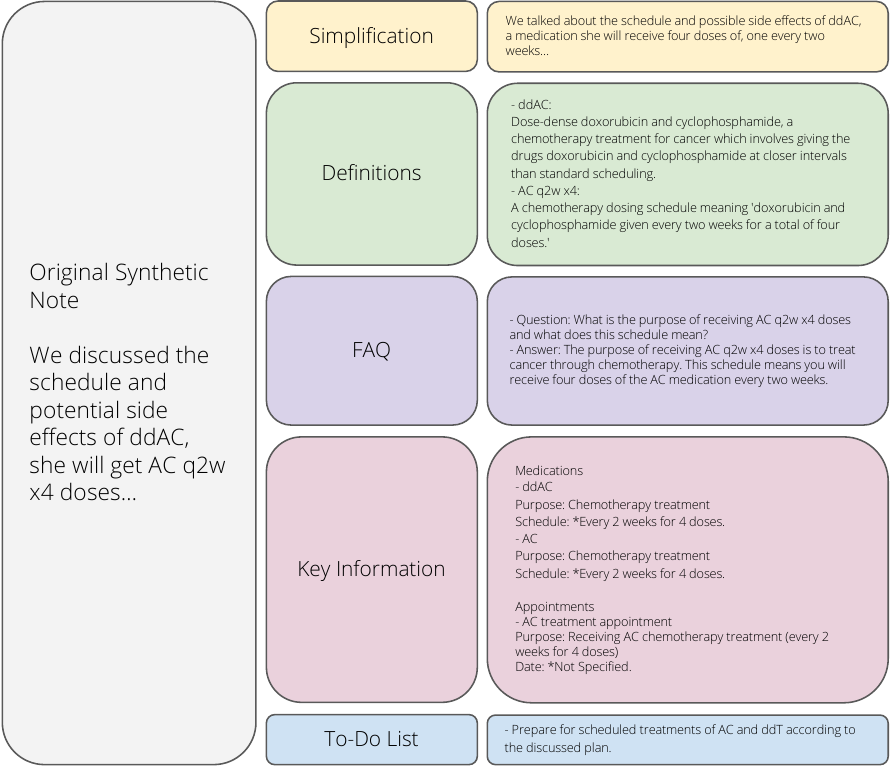}
    \caption{Examples of the augmentations considered in our interface using a sample excerpt from a synthetic clinical note. This includes the augmentations of simplification, definitions, frequently asked questions, key information, and to-do list.}

    \label{fig:examples}
\end{figure}

We developed a user interface, implemented as a web app, to integrate the augmentations presented in Section \ref{subsec:augmentations} (Figure \ref{fig:intro_ui}). The interface consists of three panels, each containing different augmentations. In the first panel, the reader has access to a tab containing the original clinical note and a second tab containing the generated simplified note. On the original note, each medical term is highlighted in bold green and underlined to indicate that the user can click on it. When a user clicks on a medical term, the page will automatically scroll to the corresponding definition which is available in the third panel. The second panel consists of three tabs for the augmentations \textit{FAQ}, \textit{Key Information}, and the \textit{To-Do List}. For \textit{FAQ}, the reader can expand on each question to see its answer. In Figure \ref{fig:examples}, we show explicit examples of the augmentations generated for one of the clinical notes used in our work.

\begin{figure}[h]
    \centering
    \includegraphics[width=\textwidth]{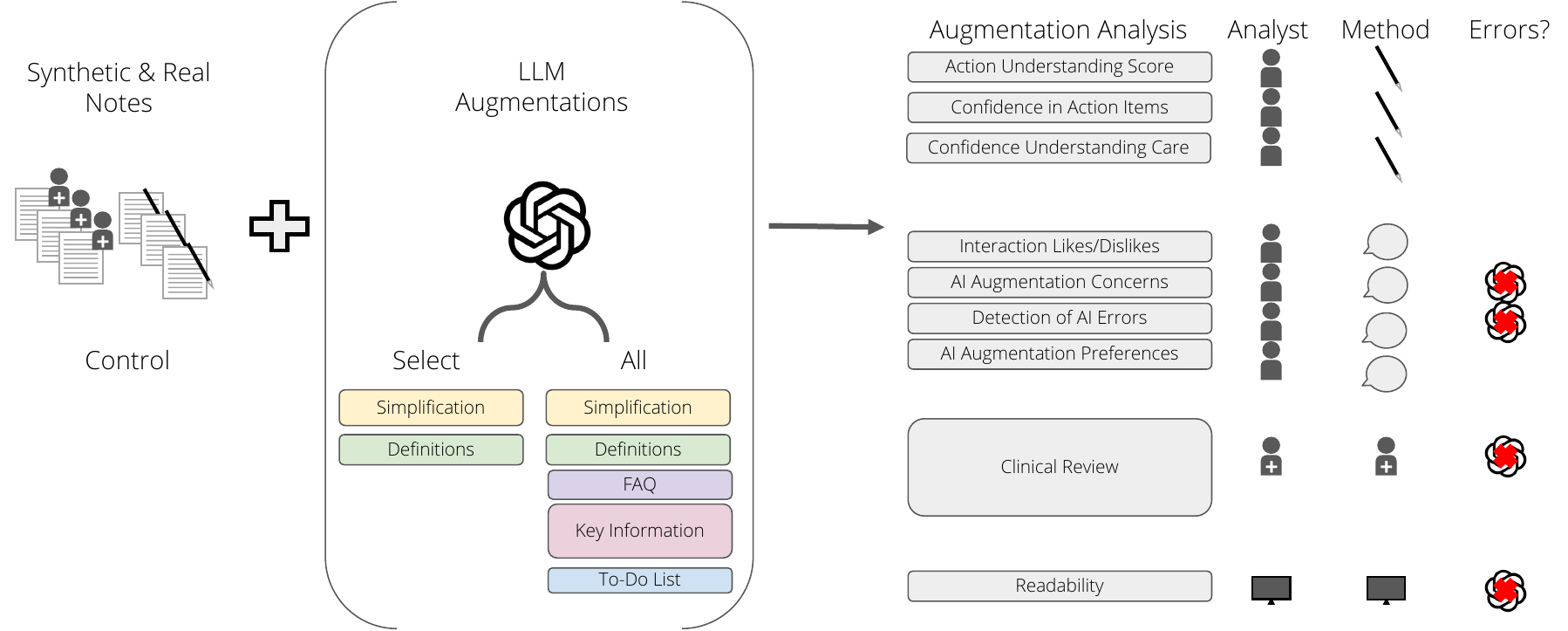}
    \caption{Overview of mixed methods evaluation of LLM augmentations for clinical notes in our work. This evaluation was conducted based on both synthetic and real notes (left of figure), and two sets of augmentations were studied: select and all. \textit{Select} included simplification and definition augmentations, and \textit{All} included these as well as \textit{FAQ}, \textit{Key Information}, and \textit{To-Do List} augmentations. Augmentation evaluation was conducted using mixed methods  (see list under Augmentation Analysis). Researchers on the team analyzed the results of the first several methods, and clinical research partners focused on clinical review (person with + icon). Readability was analyzed automatically. We used methods including both surveys  (pen icon) and interviews (chat bubble icon), as well as clinical review (person with + icon) and automated analyses (computer monitor icon). Several of these evaluations focused on identifying errors (those highlighted in last column).}

    \label{fig:overview}
\end{figure}

\section{Experimental Setup}

\subsection{Clinical Notes Collection}

To evaluate our interface, we needed a set of clinical notes for experimentation. We relied on two sources of clinical notes: synthetic notes written by a clinical research partner  and real notes donated by patients with a history of breast cancer.
The clinical research partner created four synthetic notes that mirror real clinical notes. These notes were inspired by the
 clinician research partner’s professional experience, but do not contain information about
 any of their patients. The difficulty of a clinical note is related to several factors, such as the extent of abbreviation usage and whether note-taking style sentences are used instead of full grammatical sentences. These synthetic notes were written at an intermediate difficulty,
 in the style of a clinician who is not writing for the purpose of patient comprehension.

We obtained eight real notes that were de-identified and donated by interview study participants (recruitment details can be found in Section \ref{subsec:interview_study}). The participants were self-identified patients with a history of breast cancer. Participants were told to look for clinical notes
 containing free text describing an appointment during their breast cancer treatment. Suitable notes might contain sections labeled “Impression”, “Assessment \& Plan” or “Plan”. We received written consent from participants
 before they donated their clinical notes and additional written consent before participating
 in the video interview. All participants were given a verbal overview of the purpose and risks
 listed in the consent forms, and were given an opportunity to ask questions to the research
 team before signing.  This data collection process was approved by our Institutional Review Board (IRB).

In total, our analysis drew from 12 clinical notes, each containing an average of 322 words. 
We focused on this limited set of notes because the clinician research partner determined the corpus to represent a wide variety of clinical scenarios and note-writing styles, and because of the intensive time required to curate synthetic notes and collect and de-identify donated notes. 

\subsection{Survey Study Design}\label{subsec:survey_study}

To test these augmentations' effects on patient understanding, we first conducted a randomized web-based survey with healthy participants. We recruited participants using the Prolific crowdsourcing platform with the following inclusion criteria: female-identifying, over 40 years of age, fluent in English.
Participants first completed a survey about their demographics and a health literacy test (Confident with Forms  \cite{chew2008validation}), 
then viewed three notes. Study participants were randomized to either observe synthetic or real (donated, de-identified) notes. Notes were randomly selected from a pool of four synthetic and six real notes.

Participants read the note with a given level of LLM augmentations,  answered four multiple-choice action-oriented content questions, and then self-reported their confidence and comprehension of their reading. Each note was accompanied by one of three levels of augmentations: \textit{Control} (no augmentations), \textit{Select} (only \textit{simplified} and \textit{definitions} augmentations), and \textit{All} (all \textit{simplified}, \textit{definitions}, \textit{FAQ}, \textit{key information}, and \textit{to-do list} augmentations). All participants saw all three levels of augmentations in a randomized order.  

Per note, augmentation, and participant, we obtained an action understanding score in the range $[0,1]$, a self-reported confidence score about understanding the care of the patient in $[0,3]$ (self-comprehension),  and a self-reported confidence score about understanding the next steps of the patient in $[0,3]$ (self-confidence). At the conclusion of the study, we requested a list of augmentations they would like to use again, and participants 
additionally described what they used to answer the multiple-choice questions in a free-form text response.
A clinician research partner reviewed all augmentations to check for false or misleading information.
Participants were shown a disclaimer at the start of the study, warning that some content was generated using LLMs. Information identified as misleading or false was explicitly listed and corrected at the end of the study to avoid misleading participants. 
The study took approximately 20 minutes and participants were compensated with \$3.75. Full details of the survey study can be found in Appendix \ref{apx:web_survey}. This study was determined exempt from IRB approval by our institutional IRB.

\subsection{Interview Study Design }\label{subsec:interview_study}

To evaluate the impact of our proposed clinical note augmentations and supplement the survey with more detailed feedback, we also conducted in-depth interviews to explore patient perspectives. We recruited participants who self-identified as patients with a history of breast cancer by advertising via the crowdsourcing platform Prolific and a clinical center at our institution. Details about participants' backgrounds is reported in Table \ref{tab:demographics}.
Interview studies were conducted virtually with two study team members present, including one clinical partner, following a semi-structured interview protocol consisting of exploring our proposed augmentations interface and answering a series of questions about the augmentations, including a Likert rating for each augmentation. 

One researcher conducted all of the interviews virtually. A clinical research partner was present for all interviews to answer general safety-related medical questions as needed. All sessions lasted one hour and were audio- and video-recorded via a teleconference platform, then transcribed manually. Both researchers introduced themselves at the start of the interview and reviewed the study goals. Participants were asked to answer demographic questions. A semi-structured interview protocol was developed based on the initial survey findings. Participants were asked about their prior experience reading clinical notes, and then were given a customized link to the user interface where one of their donated clinical notes and its augmentations had been preloaded. They were asked to think aloud as they used the tool, and specifically were asked what they liked and disliked, whether anything about the augmentation worried them and if they identified any errors, and finally to rate on a Likert scale from 1-5 how much the augmentation assisted their reading experience. Full details of the interview protocol and questions can be found in Appendix \ref{interview_qs}. This study  was approved by our IRB. 

Two researchers used DeDoose software and Braun and Clarke’s methods for thematic analysis for qualitative analysis of the transcribed interviews \cite{braun2006using}. One researcher first analyzed the interviews to derive major themes for an initial codebook, which was reviewed by the second researcher. Both researchers independently coded all of the interviews and iteratively updated the codebook and reconciled discrepant codes until consensus was reached.

\subsection{Error Analysis of Augmentations}\label{sec:error_method}

All augmentations underwent manual review by three clinician research partners to identify misleading or false information. One clinician research partner conducted an initial examination of all notes and augmentations, documenting any errors found along with their rationale for categorizing them as errors. The other two clinicians then validated these findings, to ensure consistency in our findings.
Once evaluated by all three clinician research partners, they were sorted into error categories (``Omission'', ``Not Simplified'', ``Incorrect Reasoning'', ``Incorrect Fact'', ``Misleading Phrasing'', ``Awkward Phrasing'', ``Hallucination'', ``Timing'', ``Brand Name'', ``Unnecessary'', ``Clinician Preference'', ``Assumed Context'') by multiple coauthors, including the clinician research partners and non-clinical research partners. We worked together to reach consensus assigning categories to each error found. The full categories are described in Appendix \ref{table:error_types}. 

In addition to manual review, we also conducted an automated analysis of the augmentations. We reported the 
Flesch-Kincaid Grade \cite{flesch1948, kincaid1975} and the percentage of words in a dictionary (Carnegie Mellon Pronouncing Dictionary Corpus Reader\footnote{\url{https://www.nltk.org/_modules/nltk/corpus/reader/cmudict.html}}), to compare the acronyms and jargon in the original versus augmented notes. 

\subsection{Statistical Analysis}
To obtain 95\% confidence intervals for mean estimates of outcomes, we used a method to correct for inter-subject variability \cite{morey2008confidence}. To compare outcome means between two conditions, we used the pairwise Wilcoxon Signed Rank test to obtain p-values. We corrected for multiple hypothesis testing with the Benjamin-Hochberg adjustment procedure with $\alpha=0.05$. We also conducted subgroup analyses, such as for those who never accessed a health portal, cancer patients and health professionals, those with an associate degree or less, and those who were not White race, to see whether overall trends held. All analyses were performed in R version 4.1.1 with the stats package.

%% file: files/results.tex
\section{Results}\label{sec:results}
\subsection{Study Population}

For our survey study, we recruited 200 participants through Prolific, many of whom had accessed a health portal, had adequate health literacy, and agreed that AI can have a beneficial impact in medicine. Most participants in both components of the study identified as female, and the majority of participants identified as white and non-Hispanic, Latino, or Spanish origin. The full demographic breakdown of the participants is in Table \ref{tab:demographics}.

For our interview study, all seven participants had accessed a portal for patients previously, demonstrated adequate health literacy, and all but one participant agreed with the statement that AI can have a beneficial impact in medicine. 

\begin{table}
\caption{Mixed methods study population breakdown by demographics, experience with patient portals and notes, health literacy, and artificial intelligence sentiment.}
\centering
\begin{tabular}{ p{40mm}  p{45mm}  p{20mm}  p{20mm}  }
\toprule
 &  & Survey (N=200) & Interview (N=7) \\
\hline
\multirow{5}{40mm}{Highest Schooling} & 8th grade or less & 3 & 0 \\

 & High school graduate or GED & 55 & 2 \\
 & 2-year college degree graduate & 35 & 0 \\
 & 4-year college degree graduate & 80 & 3 \\
 & Masters or doctoral degree graduate & 27 & 2 \\
\hline
\multirow{6}{40mm}{Race} & White & 158 & 6 \\

 & Black/African American & 24 & 0 \\
 & Asian & 5 & 1 \\
 & Native American/Alaska Native & 1 & 0 \\
 & Other & 4 & 0 \\
 & More than 1 & 7 & 0 \\
\hline
\multirow{2}{40mm}{Hispanic, Latino, or Spanish origin} & Yes & 13 & 0 \\

 & No & 187 & 7 \\
\hline
\multirow{3}{40mm}{Gender} & Female & 196 & 7 \\

 & Male & 3 & 0 \\
 & Other & 1 & 0 \\
\hline
\multirow{5}{40mm}{Age} & 40-45 & 42 & 0 \\

 & 46-55 & 68 & 3 \\
 & 56-65 & 61 & 3 \\
 & 66-75 & 28 & 1 \\
 & $>$76 & 1 & 0 \\
\hline
\multirow{6}{40mm}{Employment Status} & Employed for wages & 81 & 2 \\

 & Self-employed & 50 & 1 \\
 & Homemaker & 10 & 0 \\
 & Unemployed & 14 & 1 \\
 & Retired & 30 & 1 \\
 & Unable to work & 15 & 2 \\
\hline
\multirow{2}{40mm}{Accessed Portal for Notes?} & Yes & 122 & 7 \\

 & No & 78 & 0 \\
\hline
\multirow{5}{40mm}{\# Notes Last Year} & None & 24 & 0 \\

 & One & 25 & 1 \\
 & 2 or 3 & 41 & 2 \\
 & 4 or more & 26 & 4 \\
 & Don't know & 6 & 0 \\
\hline
\multirow{2}{40mm}{General Health Literacy \cite{chew2008validation}} & Adequate & 167 & 7 \\

 & Inadequate & 33 & 0 \\
    \hline
\multirow{5}{40mm}{“Artificial Intelligence can have a beneficial impact in medicine”} & Strongly Agree & 57 & 2 \\

 & Agree & 100 & 4 \\
 & Neither agree nor disagree & 32 & 1 \\
 & Disagree & 4 & 0 \\
 & Strongly Disagree & 7 & 0 \\ \bottomrule
\end{tabular}

\label{tab:demographics}
\end{table}

\subsection{Error Analysis}\label{sec:error}

\begin{figure}
    \centering
    \includegraphics[width=\textwidth]{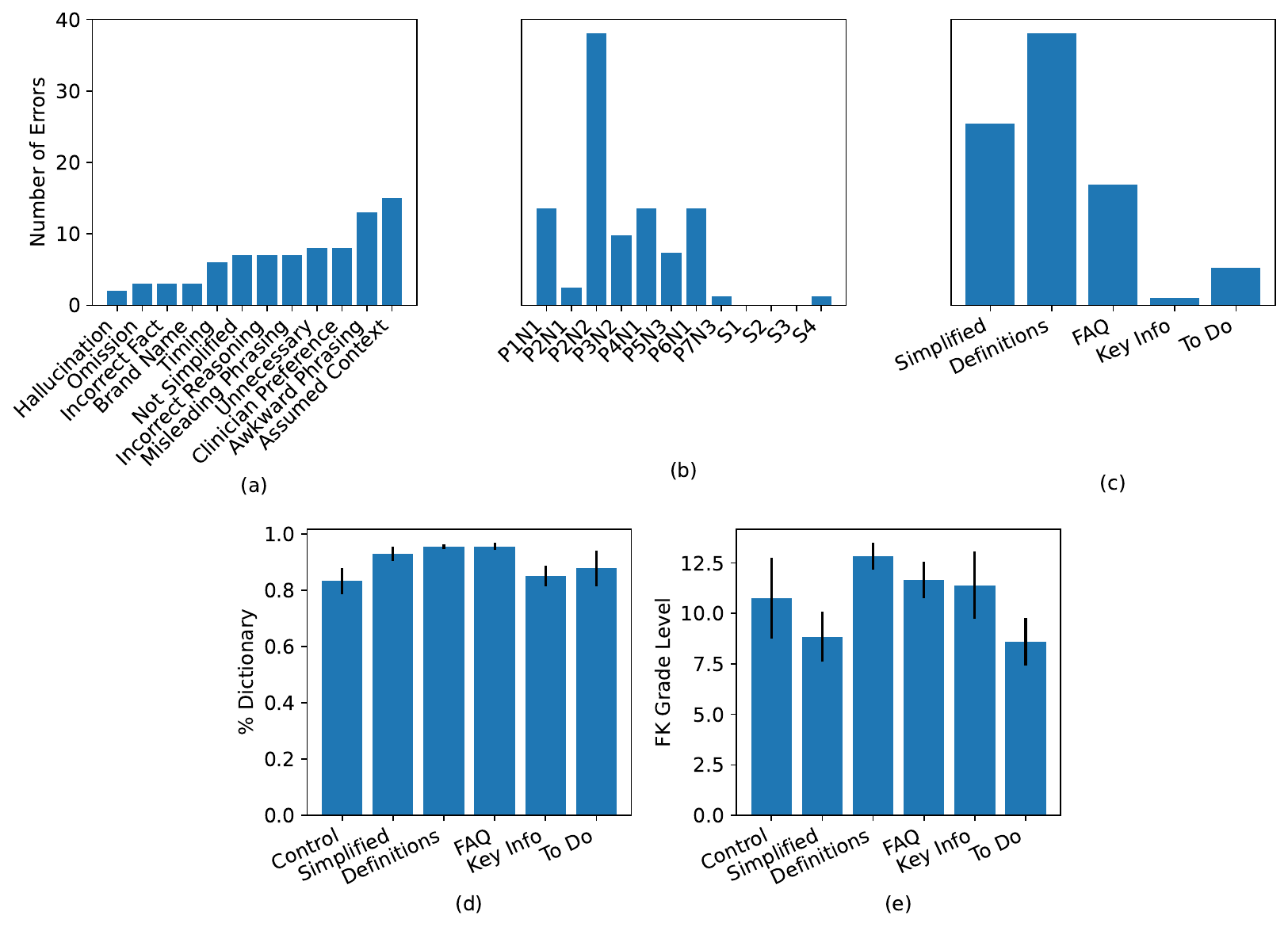}
    \caption{Error analysis of GPT-4 augmentations. (a) (b) and (c) show the number of errors in different categories, as determined by clinical reviewers. In (a), we have types of errors, such as hallucinations. The most frequent error was when GPT-4 assumed a certain context, e.g., assuming that a definition is specific to breast cancer when it truly is not.  In (b), this is grouped by augmented clinical notes, where P represents a participant's donated note, and S represents a synthetic note. There are very few errors for synthetic notes and a great deal of variation in the number of errors in donated clinical notes. In (c), the number of errors is shown in each of the augmentations, with definitions containing the most errors. (d) and (e) are the result of automated analysis of GPT-4 augmentations. The error bars on subfigures   (d) and (e) represent confidence intervals computed using t-scores, note that we omit from the analysis empty note augmentations.  (d) contains the percentage of words that are found in the Carnegie Mellon Pronouncing Dictionary Corpus Reader.  Control has the least, as there are potentially more acronyms and jargon. (e) shows the Flesch Kincaid grade level score, which is generally comparable or less than control in the augmentations, except for definitions. From interviews, we know that definitions still do contain jargon that can be confusing.}

    \label{fig:error}
\end{figure}

We report the results of the error analysis in Figure \ref{fig:error}. We found that GPT-4 can make misleading errors in the context of breast cancer clinical note augmentations, and identified 82 total errors. In total, 28 (34\%) errors were considered to be potentially harmful and misleading (Figure \ref{fig:error} (a)). 
However, minor errors were most common, especially assumed context (Figure \ref{fig:error} (a)). For example, a definition augmentation specifying that lower outer quadrant referred specifically to the breast, though the term can refer to a specific part of other tissues as well. Many of these errors appeared in the \textit{definitions} augmentation, and errors were especially common in one note's augmentations (Fig. \ref{fig:error} (b-c)). 
In particular, most errors occurred in augmentations to real notes, while only one error appeared in augmentations to the synthetic notes (Figure \ref{fig:error} (b)). 

In the automated analysis, we found that the percentage of words in the dictionary increased over \textit{control} in all augmentations. Notably, when comparing \textit{control} versus the \textit{simplified} augmentation, the percentage increased for all notes, and the results of a paired-sample t-test are  ($t=-4.61$, $p=0.0007$)
(Figure \ref{fig:error} (d)).
Moreover, the Flesch-Kincaid grade level decreased compared to \textit{control} for the \textit{simplified} and \textit{to-do list} augmentations, when comparing \textit{control} versus the \textit{simplified} augmentation, the results of a paired-sample t-test are ($t=2.24$, $p=0.047$). However, augmentations like \textit{definitions} increased in grade level (Figure \ref{fig:error} (e)), which could be due to additional length and medical terminology in each of the definitions. This was also observed by some of the qualitative interview participants.  Future work could explore further simplification of the \textit{definitions} and evaluation on a larger number of notes.

\subsection{Quantitative Analysis}

\begin{figure}
    \centering
    \includegraphics[width=0.8\textwidth]{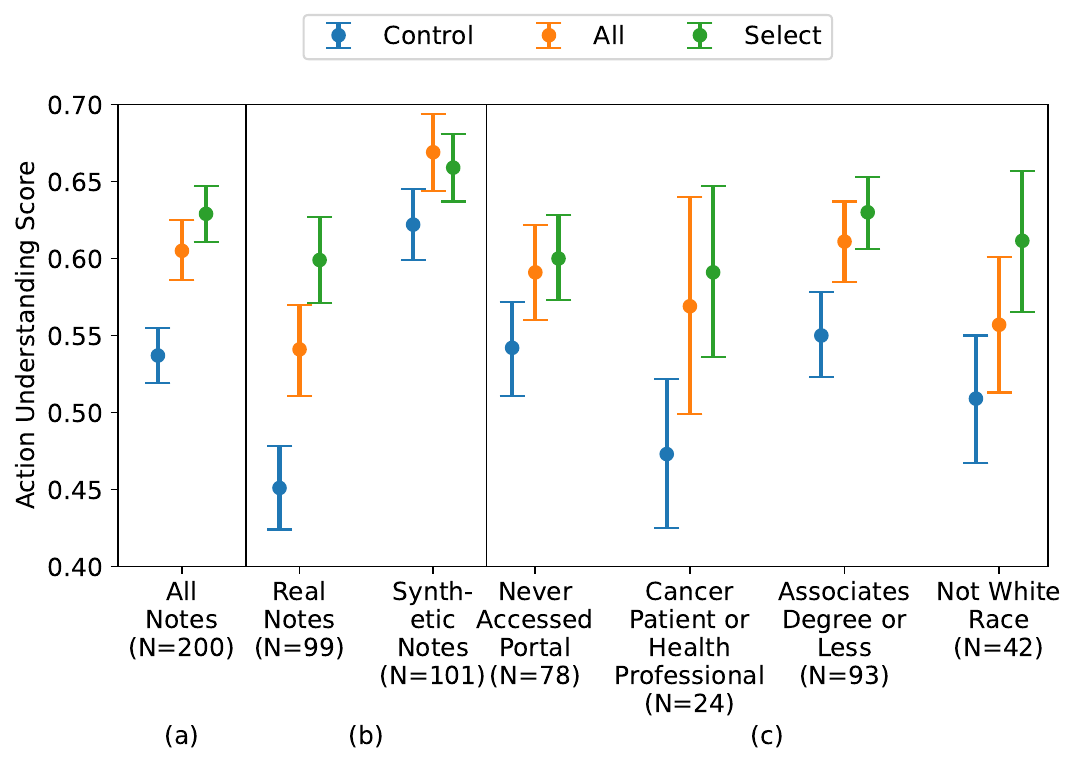}
    \caption{Results from our web-based survey for participants' action understanding score for the Control, All, and Select conditions. We show the scores for all notes (a), synthetic and real notes (b), and scores across multiple subgroups (c). Generally, Select (definitions and simplification augmentations) leads to the best action understanding score and improvement over control. We note that both the All and Select conditions significantly increase action understanding score on the aggregate.}

    \label{fig:quant}
\end{figure}

In the survey study (N=200), we found that participants reading notes with either \textit{Select} or \textit{All augmentations} had statistically significantly higher action understanding scores, self-reported comprehension scores, and self-reported confidence scores when considering both real and synthetic note conditions in aggregate. 
Notably, with \textit{All augmentations} compared to  \textit{Control} (no augmentations),  action understanding scores increased from 0.537 to 0.605 (p=0.02), self-comprehension scores increased from 1.45 to 2.25 (p$<$ 0.001), and self-confidence scores increased from 2.17 to 2.95 (p$<$ 0.001) (Figure \ref{fig:quant} (a)). We found similar observations when comparing \textit{Select} to \textit{Control}. 

We found that the trend of \textit{Select} and \textit{All augmentations} increasing self-reported and action understanding scores held when looking at specific subgroups of the study population, which we selected prior to analysis of the results (Figure \ref{fig:quant} (c)).  
We note that \textit{Select} and \textit{All augmentations} had higher action understanding, self-comprehension, and self-confidence scores on average in each of the four subgroups, though we did not evaluate statistical significance. We also looked at other possible subgroup categorizations post-hoc and observed similar trends. 

When we split the participants based on the condition assigned of observing either synthetic notes or real notes, we found that \textit{Select} and \textit{All augmentations} statistically significantly improved action understanding scores only for the real notes and not for the synthetic notes (Figure \ref{fig:quant} (b)). For participants in the real notes condition,  action understanding scores increased from 0.451 in \textit{Control} to 0.541 in \textit{All augmentations}  (p=0.04); for participants in the synthetic notes condition, scores changed from 0.622 in \textit{Control} to 0.669 in \textit{All} (p=0.30). 

The top two modifications that participants indicated they would use if they were to read another clinical note were \textit{simplifications} chosen by 83.5\% (95\% CI 77.7-88.0) of participants and \textit{definitions} chosen by 72.5\% (95\% CI 65.9-78.2) of participants. The remaining preferred modifications in order are \textit{To-Do-List} 42.0\% (95\% CI 35.4-48.9), \textit{FAQ} 41.0\% (95\% CI 34.4-47.9), and \textit{Key Information} 35.5\% (95\% CI 29.2-42.3). In the answers to the question ``What did you use to answer these questions? Was it easy or hard?”, the most common word was "simplified" (occurred 97 times), followed by "notes" (71), "easy" (58), "original" (54), and "definitions" (54). This highlights that \textit{simplified} and \textit{definitions} augmentations were most useful to participants.

\subsection{Qualitative Analysis}

\begin{table}[h]

\caption{Common themes from the interview study organized into three categories: Current practices and attitudes towards note reading, Wins with the modifications, and Concerns with the modifications.  }
\centering

\begin{tabular}{  p{50mm}  p{90mm}   }
\hline
\multicolumn{2}{l}{\textbf{Current practices and attitudes towards note reading}} \\
\hline
Reading notes to gather information & “I wanted to know the dates when that was diagnosed, and maybe when was the last time I did a test” [P7]
\medskip

“Figure out when my appointments are. I'll look at blood work, you know, if anything's weird.” [P5] \\
\hline
Reading notes to check the narrative & “[To find out] what my doc actually thought as opposed to what he said to me” [P4] \medskip

“You kind of want that insight into like what the doctor's take was on it” [P2] \\
\hline
Medical terminology makes notes difficult to understand & “Usually, I have to Google some words. Especially in the beginning when you’re first diagnosed, there’s a lot of initials and words that a lot of people don’t understand” [P1] \medskip

“At the time, I looked them up but they’re hard to retain even. It’s just I don’t use them in everyday language” [P2] \medskip

“I have to do a lot of Googling though to see what this means and what that means and whether it’s significant or not.” [P3] \\
\hline
\multicolumn{2}{l}{\textbf{Wins with the modifications}} \\
\hline
Improved understanding of the note & “It gives me more comfort and information going, oh, that’s what they mean” [P6] \medskip

“It gives you all the information and it’s actually less stressful to read. I think because it’s not as overwhelming.” [P2] \medskip

“It’s definitely easier to understand and at least one of the questions I had, I got answered here” [P5] \\
\hline
Ease of use & “This is like a personalized Wikipedia page for my diagnosis” [P3] \medskip

“It’s just nice to have it all there if I want it. I don’t have to look up everything but I can look up what I want to.” [P2] \\
\hline
\multicolumn{2}{l}{\textbf{Concerns with the modifications}} \\
\hline
Missing or misleading information & “I feel like this makes assumptions that can’t really be made” [P5] \medskip

“I guess I feel like blood work should be in there.” [P6] \\
\hline
More information can be confusing & “It made me feel a little bit more worried because they were things that I hadn’t conceived of yet” [P2] \medskip

“That's a bit overly heavy on medical jargon.” [P3] \medskip

“If I'm going to have a medication list, I think I would want the medication list of what I'm on or what I will be having, not everything that came up.” [P5] \\
\hline

\end{tabular}

\label{tab:qual}
\end{table}
\clearpage

Interview participants (N=7) described three major themes around note-reading (Table \ref{tab:qual}):  (1) gathering information, (2) checking the narrative, and (3) understanding medical terminology.  
Participants shared that they read notes for facts such as lab results and future appointments, and to help their own memory and ensure they understood the visit correctly. Participants reported that they read notes in order to assess the narrative, for example ensuring that the clinician understood them and was aligned in their priorities. Many also described notes as a useful window into the clinician’s thinking: “You kind of want to have that insight into what the doctor’s take was on it” [P2]. All participants reported that medical terminology makes notes hard to understand, which can lead to time-consuming research and fear of the unknown.


All participants rated the \textit{definitions} augmentation highest in the Likert ratings in terms of greatest benefit to note-reading. Participants described two main benefits with augmentations: (1) improved understanding of the note and (2) ease of use (Table \ref{tab:qual}). Participants reported that augmentations made notes easier to read, less stressful, more personable, and in some cases prompted new questions and ideas. A few participants commented that the tool performed intuitively, highlighting jargon they would have struggled with and emphasizing takeaways they also found to be important. Many reported that it was convenient to have the tool integrated with the note and that the tool kept them focused: “When I didn't have this, I would go off on websites that probably didn't even apply to my type of cancer and they were doom and gloom, I would literally sit at my desktop and cry. It was so scary” [P4]. Participants also liked that they were able to toggle between the original note and the augmented note in order to check for accuracy.


Two major themes emerged around concerns: (1) That the augmentations may miss something or mislead patients and (2) more information could be confusing (Table \ref{tab:qual}). One participant said, ``I feel like there’s a tiny bit of anxiety being like, 'I'm having to trust that the AI got it right and that this is correct''' [P2]. Others reported that augmentations may lose nuance in their oversimplification, misrepresent the gravity of the situation, or cause patients to think they understand more than they do. Participants also pointed out perceived discrepancies, such as a medication in the generated \textit{to-do list} that they were no longer taking. We present our full codebook for the interviews in Appendix \ref{apx:codebook}.

%% file: files/discussion.tex
\section{Discussion}

In this mixed methods study, including manual and automated error analysis, quantitative surveys, and qualitative interviews, we demonstrated that an LLM can be used to improve understanding of clinical notes. Participants performed better on action-oriented content questions and reported self-perceived improvements in understanding and confidence when reading notes with LLM-generated augmentations. 

Subgroup analysis of vulnerable populations—less educated and nonwhite participants—showed that the benefits persist. These groups have previously reported the most benefit from reading clinical notes \cite{gerard2018importance}, and leveraging LLMs to improve note readability may be an important tool for organizations’ diversity, equity, and inclusion efforts.  In our qualitative evaluation, patients reported that different people prefer different amounts of information about their health, and those information needs can vary over time. 

Through clinical review of LLM augmentations, we found and characterized error modes, making it clearer what work is needed to improve the accuracy of LLMs before they are deployed as standalone patient-facing tools. 
First, errors were more common in augmentations for real donated notes compared to augmentations for the carefully-written synthetic notes, suggesting that a real-world deployment would still be impacted by the quality of the original note. 
Furthermore, while hallucinations were more rare, patients' trust in qualitative interviews was still impacted by the errors we identified, even for those who were enthusiastic about the use of AI in medicine. We note that in our web-based survey, we did not find an impact in action understanding score due to errors, likely because the identified errors only impacted the ability to answer the action-oriented questions in one note, and participants could still view the original note.  Especially in the presence of errors, carefully presenting LLM-generated advice is extremely important to study for future deployment  \cite{ghassemi2023presentation}. In this study, we began to consider this by presenting both the original and augmented notes for patients to reference.

Relatedly, note-based augmentations can only be as accurate as the underlying note, and LLMs may amplify existing documentation errors and biases \cite{bell2020frequency,adam2022write}. 
In our study, patients preferred augmentations that provided medical definitions and simplified existing documentation. In the interviews, several participants also described the value of viewing the original note and emphasized the importance of reading notes not only for factual information but also to understand the clinician’s perspective and check for alignment. In the survey, we similarly hypothesize that \textit{Select} augmentations slightly outperformed \textit{All} augmentations because the additional augmentations in \textit{All} may have opened up more potential misunderstandings.

There are limitations to this study. We used a mix of real and synthetic notes in order to expand our pool of available notes for augmentation, but the synthetic notes were likely less complex and variable than the real donated notes. Augmentations led to improved scores on the real notes only, and most of the identified errors were found in the real notes. 
Future experiments may seek to evaluate the patient experience prospectively on real notes only. Evaluating a larger, more diverse cohort of patients at a medical center may also help to minimize the selection bias present in our study; participants who volunteered for our study may have been more likely to view AI favorably. In addition, further research is needed to improve LLMs for clinical use, particularly for tools that are used by patients; additional fine-tuning to reduce errors \cite{liu2023leveraging}, may help create tools that truly empower patients. However, given that errors may never be fully mitigated, another important direction to explore is improved interface design to make it easier for patients to notice potentially misleading facts in-line. Additional prompt engineering may help create custom augmentations that are tailored to patient readiness, their education level, or even their preferred language \cite{zhu2023multilingual}. Further research is needed to understand how clinicians' and patients' voices can be preserved while alleviating documentation and improving communication and education.

\section{Conclusion}
In this work, we presented an end-to-end tool that improves information accessibility by giving patients the ability to view their clinical notes with five different LLM-enabled augmentations. We used the tool to study the effect of these augmentations both qualitatively and quantitatively, aiming to understand both patients' attitudes towards AI and their understanding of actionable health information. We were also able to understand patients' prioritization and preferences for choices of augmentation alongside clinician ratings of errors across these same augmentations. This characterization can inform future design choices in information accessibility, across domains.

Large language models hold tremendous promise in their potential to transform healthcare and empower patients with their own data. Careful design and implementation could provide the opportunity to change the patient-clinician relationship for the better as we move into this new era of health technology. However, thoughtful deployment and continued participatory design is required to modulate trust between users and AI, particularly in high-stakes scenarios like healthcare.

%% file: appendix.tex
\clearpage
\appendix

\section{User Interface Overview}

\begin{figure}[h]
    \centering
    \includegraphics[width=\textwidth]{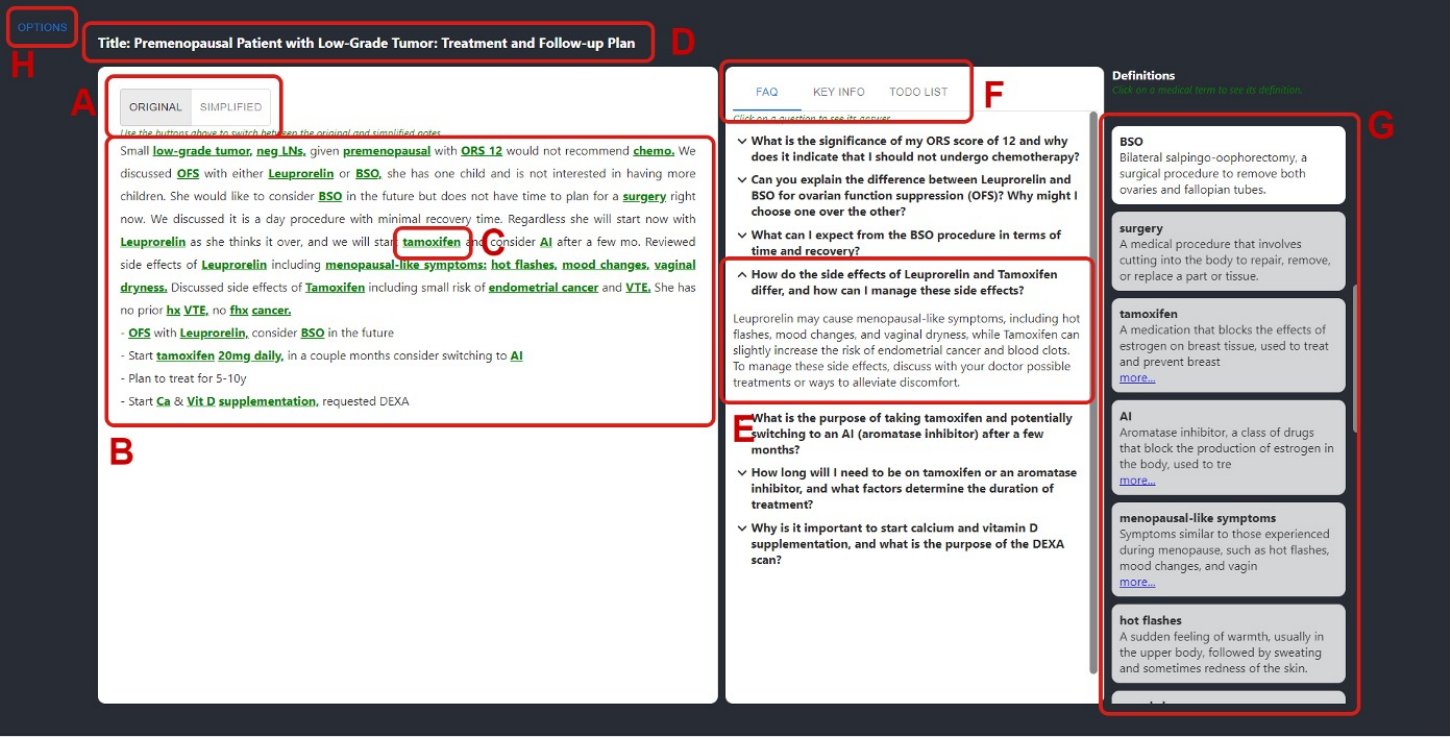}
    \caption{
    Direct screenshot of user interface for augmented clinical notes, with annotations: 
    \textbf{(A)} The user is able to switch between their original note and a simplified version using tabs.
    \textbf{(B)} The user's note (either the original or simplified depending on the selected option) is displayed.
    \textbf{(C)} Medical terms are highlighted in bold green and underlined to indicate that the user is able to click on them.
    When a user clicks on a medical term, the page will automatically scroll to the corresponding definition in \textit{(G)}
    \textbf{(D)} A title briefly describes the content of the clinical note.
    \textbf{(E)} Users can explore a series of questions about the note. 
    By clicking on a question, its corresponding answer will be revealed.
    \textbf{(F)} Tabs enable navigation between different augmentations. 
    The figure does not depict the \textit{Key Info} and \textit{To-Do List} augmentations, which display medication and appointment data as well as action items, respectively.
    \textbf{(G)} A scrollable sidebar is utilized to present definitions of medical terms.
    \textbf{(H)} An options menu allows the user to toggle the augmentations they wish to enable for their note.
    }
    \label{fig:ui_full}
\end{figure}

\begin{figure}[h]
    \centering
    \includegraphics[width=\textwidth]{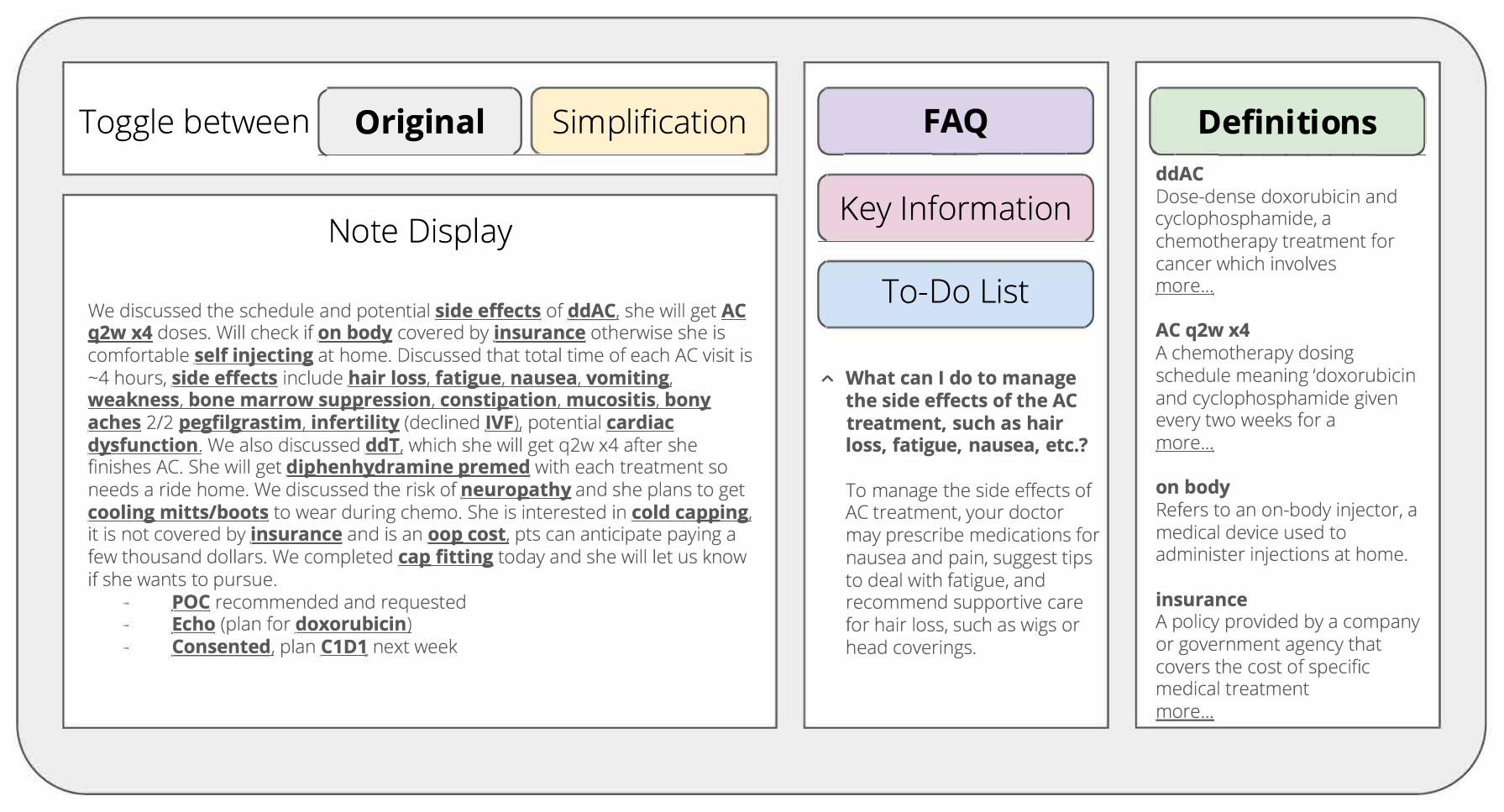}
    \caption{User interface concept that aligns with Fig. \ref{fig:examples}.}
    \label{fig:ui}
\end{figure}

\clearpage

\section{Interview Questions} \label{interview_qs}
The researcher began by asking the participant questions about their experience reading clinical notes, as follows, including the health literacy question in bold:
\begin{itemize}
    \item \textit{``Have you previously accessed a patient portal to read clinical notes?''}
    \item If yes, \textit{``About how many clinical notes have you read in the past 12 months?''}
    \item \textit{``What are your goals when reading your clinical notes?''}
    \item \textit{``Would you say that you usually achieve these goals?''}
    \item \textit{``What challenges do you face when you read your clinical notes?''}
    \item \textit{``Have you previously accessed a patient portal to read clinical notes?''}
    \item \textit{``In the last 12 months, how many visits have you had at any healthcare facility?''}
    \item \textbf{\textit{``How confident are you at filling out medical forms by yourself?''}} (Not at all, A little bit, Somewhat, Quite a bit, Extremely) \cite{chew2008validation}
    \item `\textit{`How much do you agree with the following statement: `Artificial intelligence can have a beneficial impact in medicine.'''} (Strongly Agree, Somewhat Agree, Neither Agree Nor Disagree, Somewhat Disagree, Strongly Disagree)
\end{itemize}

Next, the researcher shared a custom link with the participant for them to access the user interface  and view the clinical note they previously submitted.
The researcher prompted the participant to refamiliarize themselves with the original note and to share any thoughts they have out loud. 
After about five minutes, the researcher asked the following questions
\begin{itemize}
    \item \textit{``Is there anything you found hard to understand?''}
    \item \textit{``Other than that, are there any questions you have about the contents of the note?''}
\end{itemize}
The participant was then directed to the options menu, where they could enable and read descriptions of each augmentation.
The participant was prompted to read about each augmentation and enable the one that most interested them.
The participant was then asked to explore their clinical note with the active augmentation and to share their thoughts out loud.
The researcher then asked the following questions

\begin{itemize}
    \item \textit{``What do you like about this augmentation?''}
    \item \textit{``What do you dislike about this augmentation?''} 
    \item \textit{``Is there anything that worries you about this augmentation?''}
    \item \textit{``Do you notice any mistakes that this augmentation made with regard to your treatment or diagnosis?''}
    \item \textit{``On a scale from one to five, how much does this augmentation add to your reading experience?''}
\end{itemize}

Then the participant was prompted to return to the \textit{Options} menu, disable the active augmentation, and repeat the process with the next augmentation they were interested in.

Once all of the augmentations were explored individually, the participant was prompted to enable as many augmentations as they would like, to explain the selection they made, and provide overall feedback on their experience reading their original and augmented notes.

\clearpage
\section{Full Codebook for Coding  Interviews}\label{apx:codebook}


\begin{longtable}{  p{50mm}  p{90mm} }
\hline
\multicolumn{2}{l}{\textbf{Why patients read notes}} \\
\hline
Accuracy; check if they got facts right & “I want to have an understanding of what’s going on and see if anything has been overlooked.” [P3] \\
\hline
Alignment

Check if doctors got the context/story right

Check if doctor/patient priorities align & “To make sure that the doctor and I had the same conversation. Did I miss anything?” [P1] \\
\hline
Memory; to remember something or see if they missed anything that was discussed & “...if I wanted to review my notes, you know, post a procedure, this is very helpful to recall what was done and what it all meant.” [P7] \\
\hline
Preparation; to review and ready questions for the visit & “Sometimes the notes will reflect the test I just had. If I get them before our meeting I like to look at them so I can ask them questions.” [P1] \\
\hline
Comprehension; to check your own understanding & “I like to understand …the definitions of what I'm reading.” [P7] \\
\hline
Information-seeking

I consider myself detail-oriented and like more information

More information can be comforting & “Sometimes research is how I self-soothe. The more information I have, it makes me feel more prepared and equipped to deal with any scenario.” [P2] \\
\hline
Space; to digest the visit alone and at their own pace & “During the meeting I understood everything but when you’re in the meeting you’re taking it all in and you really don’t know which questions to ask at the time and you kind of have to think afterwards.” [P1] \\
\hline
To understand the doctor’s \textit{perspective} or thinking & “You kind of want that insight into …what the doctor's take was on it.” [P2]  \\
\hline
For health monitoring

Get results

Check appointments & “Wanting to see …trends and being able to really toggle back and forth between past …results and past things.” [P2] \\
\hline
\multicolumn{2}{l}{\textbf{Challenges patients face when reading notes}} \\
\hline
Medical terminology can make the note hard to understand & “It’s like having to translate [the note] from a different language.” [P2]“I have to do a lot of Googling though to see what this means and what that means and whether it’s significant or not.” [P3] \\
\hline
Fear due to a lack of understanding & “As I was reading it, I was wondering how bad it was. There was the term invasive. And that was scary.” [P3] \\
\hline
Patient portal usability & “My medical place uses MyChart and sometimes it’s not the friendliest to go back and forth in-between…” [P2] \\
\hline
Difficulty finding answers through self research & “I would look for the answer, which was extremely time consuming, and I didn’t always get a good answer.” [P2]“You have no idea how many papers I have…I wouldn’t have to flip through everything just to look for [one] definition or one appointment.” [P4] \\
\hline
\multicolumn{2}{l}{\textbf{Wins}} \\
\hline
Translating medical terminology helps with understanding & “If I want to do more research I can but now I have a little snapshot of what they’re talking about.” [P1] \\
\hline
Convenient to see relevant information and answers in one place & “I think it’s helpful. Instead of having to run to Google.” [P1] \\
\hline
Simplified notes can seem more personal & “Like a patient took these notes, not a machine.” [P7] \\
\hline
This augmentation makes the note easier to read

Less stressful & “[The simplified note] is actually less stressful to read. I think because it’s not as overwhelming.” [P2] \\
\hline
Prompt new ideas and questions & “It also asked questions that I wouldn’t think to ask, that I would like the information for.” [P4] \\
\hline
Keeps patients focused  & “I wouldn’t go off on all of these different websites…this doesn’t give me the worst possible scenario, I like that.” [P4] \\
\hline
Patients can do learning on their own time & “I could go home and look at them whenever, or, you know, however long it took me to understand it…” [P4] \\
\hline
Mod can help caregivers & “Cancer patients are not are not always–I couldn’t even read at that time, or pay bills–I was a hot mess. So I like the fact that the caretakers can take care of [the to do list].” [P4] \\
\hline
Augmentation helps with memory or organization & “I usually kind of look at the clinical notes just to make sure that I didn’t miss what they told me when we were in the exam.” [P6] \\
\hline
Different levels of simplification are needed at different times by different patients & “I like that it’s almost like you can increase how complicated it is as you feel ready for it.” [P2]“I would imagine there are definitely people who would want the more simplified version.” [P5] \\
\hline
Tool has intuitive performance & “I like how it phrased things, how it chose to simplify things.” [P2]“This presents some more information that I didn’t have and it’s detailed but it’s not too detailed.” [P3] \\
\hline
Being able to see the original version is important

Builds trust in the AI system

Allows the patient to compare what the doctor wrote & I would click on the simplified [note] just to get..bullet points. If I really wanted to have more detailed information, I would click on the original.” [P7] \\
\hline
\multicolumn{2}{l}{\textbf{Concerns}} \\
\hline
The augmentations may miss something important & “”I feel like there’s a tiny bit of anxiety being like, ‘I’m having to trust that the AI got it right and that this is correct.’” [P2]“If you simplify it too much, people are going to miss some things and be like, ‘Why didn’t anyone tell me this? Why didn’t anyone tell me that?’.” [P5] \\
\hline
The augmentations may lose context in their simplification & “I don’t know that it 100\% captures the doctor’s thinking in the same way.” [P5]“I know what kind of treatment I had but if someone else were to read this, what is ‘extra cancer treatment’? Was it radiation? Was it chemo?” [P1] \\
\hline
An oversimplification may mislead patients  & “I think it would make people think they understood more than they may actually understand.”  \\
\hline
Professionalism and respect for gravity of situation & “I almost feel like it’s kind of talking down to you. It reads to me like at a third grade level.” [P6]“It’s so impersonal that this feels a little casual but it is a lot easier to understand.” [P2] \\
\hline
Too much information could be concerning/worrisome & “It almost adds more questions because it’s like, what is the difference between calcifications and punctate calcifications?” [P2] \\
\hline
Introduces more complexity (e.g., definitions with jargon) & “Some words I still don’t understand in the definitions.” [P4]“I don’t understand…prophylactic antibiotics. That’s–I don’t necessarily know what that means, so I would have to go back and look it up.” [P4] \\
\hline
The use of generic versus brand names can be confusing & “The inconsistency with the naming…[I could] have gotten the wrong information from this…” [P5] \\
\hline
Mod is not useful & “It doesn’t really add anything” [P7]“I’m not a fan of a to-do list here. Just because: check off ‘continue to monitor symptoms’. You don’t want to check that off.” [P1] \\
\hline
Errors the patient identified in the mods & “I assume ‘LMP’ means last menstrual period. I don’t know why it says over here that it was 12/17 and here it says it was 11/17. That doesn’t make sense.” [P5]“It’s a little weird…we went from this [AC-T] to this [T-AC] and it’s not clear that those two things are the same thing.” [P5] \\
\hline
\multicolumn{2}{l}{\textbf{Cancer specific themes}} \\
\hline
A sense of “betrayal” & “My body had turned traitor, so they could all be rebelling and I wouldn’t know. So it was that sense of surprise and not trusting my body.” [P2] \\
\hline
Anxiety related to prognosis & “Basically the big fear was, ‘Am I going to have chemotherapy and radiation?’ and ‘Is this terminal?’ I just sort of had blinders on to, ‘Where is this headed?’” [P2] \\
\hline
Patients learn more about their disease over time & “...the earlier you are in your treatment, the more helpful this will be. I'm several years of doing this…” [P1] \\
\hline
\multicolumn{2}{l}{\textbf{Other}} \\
\hline
Suggestions for the augmentations & “I would personally like the definitions to be closer. Instead of in the middle.” [P2] \\
\hline


\end{longtable}





\clearpage
\section{Categories of Errors}\label{table:error_types}


\begin{longtable}{  p{35mm}  p{55mm}   p{45mm}  }
\hline
Type of error & Example augmentation classified as an error & Reasoning behind error \\
\hline
Hallucinated information & \textit{Common Question}

What are the pros and cons of having the mastectomy and tissue expander placement in Location0 versus Location1? Having the mastectomy and tissue expander placement in Location0 might be more convenient if you live there, but it requires follow up visits. On the other hand, doing the procedure in Location1 would allow you to return there for tissue reconstruction when needed, but may involve more logistical challenges. & Augmentation mixed up the description of Location0 versus Location1  \\
\hline
Incorrect medical reasoning & \textit{Common Question}

How successful was my bilateral mastectomy and SLNB in addressing my invasive ductal carcinoma? The bilateral mastectomy and SLNB were performed to remove the cancerous tissue and check if it has spread to nearby lymph nodes. Your report indicates that no lymph nodes were involved (N0), which is a positive outcome as it suggests that the cancer might not have spread to other areas. & N0 is a description of the stage of the breast cancer, and not a reflection of the success of the surgery. \\
\hline
Incorrect medical fact or expansion & \textit{Simplified}

The patient had surgery on their right breast to remove a type of breast cancer called invasive ductal carcinoma with some features of lobular cancer. & Breast cancer can have lobular features. This augmentation implies that there was a separate type of cancer, lobular cancer. \\
\hline
Omission in augmentation & \textit{Simplified}

The patient is a 47-year-old woman who has had treatment for a specific type of breast cancer (DCIS ER/PR negative HER2+). & The original noted the staging of the breast cancer as pT1c, which is critical information to include even in a simplified version of a note \\
\hline
Phrasing could affect patient & \textit{Definition}

vus POLE: A variant of uncertain significance (VUS) in the POLE gene, which may or may not contribute to cancer development & Recommend changing to: “A variant of uncertain significance (VUS) in the POLE gene, which indicates we do not have enough information to understand whether this genetic mutation contributes to cancer development” in order to minimize worry \\
\hline
Time/tense issue & \textit{Structured }

Repeat Echo on 11/19 & The original note stated “Echo 11/28=70\%-2/7=60\%-repeat 11/19=65\%” which indicates the 11/19 echo was already completed \\
\hline
Assumed context & \textit{Definition}

Lower outer quadrant: The lower outer section of the breast, divided into four parts called quadrants & The original note stated, ``...lower outer quadrant of the left breast...'' Lower outer quadrant can refer to a specific part of many different types of tissues, not just breast. This definition may lead patients to believe lower outer quadrant is a terminology specific to breast tissue. \\
\hline
Did not simplify & \textit{Simplified}

Get blood tests (CBCD, CMP) one year before returning & Recommend simplifying the abbreviations CBCD and CMP \\
\hline
Needs generic name & \textit{Definition}

Adjuvant TH: Adjuvant therapy given after surgery or radiation, combining taxol (a chemotherapy drug) and herceptin (targeted therapy for HER2-positive breast cancer) & Recommend providing the generic names for Taxol and Herceptin \\
\hline
Phrasing is awkward & \textit{Common Question}

What is adjuvant hormonal therapy and why do you strongly recommend it for my treatment plan? & The phrasing of “why do you” may make the reader think the clinician is specifically answering this question \\
\hline
Provider preference & \textit{Definition}

Left breast ductal carcinoma in situ

Ductal carcinoma in situ & The definition for “left breast ductal carcinoma in situ” is redundant if “ductal carcinoma in situ” has already been defined \\
\hline
Unnecessary excess information & \textit{Definition}

Nipple discharge: fluid seeping out of the nipple, which may be a symptom of a benign disorder or a sign of breast cancer & “which may be a symptom…” is unnecessary commentary on the definition \\
\hline

\end{longtable}



\clearpage

\section{Web Survey Details}\label{apx:web_survey}

\subsection*{Participants}

We recruited participants using the crowdsourcing platform Prolific \footnote{\url{https://www.prolific.com/}} with the following inclusion criteria: age 40–100, sex: female, fluent languages: English, approval rate: 98–100. The interface for the user study was implemented using React and NodeJs with a Flask server.

\subsection*{Demographics Survey}

Do you work, or have you ever worked, in healthcare?
\begin{itemize}
\item  Yes, I am a healthcare professional (Medical Doctor, Doctor of Osteopathic Medicine, Registered Nurse etc.)
\item  Yes, I was previously employed as a healthcare professional (Medical Doctor, Doctor of Osteopathic Medicine, Registered Nurse etc.)
\item  No, I have never been employed in healthcare.
\end{itemize}

Do you have a personal history of cancer?
\begin{itemize}
\item  Yes, I either currently have cancer or have had cancer at some point in the past.
\item  No, I have never been diagnosed with cancer.
\end{itemize}

Have you ever accessed a patient portal to read clinical notes?
\begin{itemize}
\item  Yes
\item  No
\end{itemize}

How many clinical notes have you read in the past 12 months?
\begin{itemize}
\item  None
\item  1
\item  2 or 3
\item  4 or more
\item  Don't know / don't remember
\end{itemize}

How do you use the portal? Select all that apply
\begin{itemize}
\item  As a patient
\item  As a care partner
\item  As a parent
\end{itemize}

In the last 12 months, how many healthcare visits have you had at any healthcare facility?
\begin{itemize}
\item  0
\item  1-5
\item  6-10
\item  11-20
\item  More than 20
\end{itemize}

In general, how would you rate your overall health?
\begin{itemize}
\item  Excellent
\item  Very good
\item  Good
\item  Fair
\item  Poor
\end{itemize}

How confident are you at filling out medical forms by yourself? (Note for appendix: health literacy question \cite{chew2008validation})
\begin{itemize}
\item  Not at all
\item  A little bit
\item  Somewhat
\item  Quite a bit
\item  Extremely
\end{itemize}

What is the highest level of schooling you have completed?
\begin{itemize}
\item  8th grade or less
\item  High school graduate or GED
\item  2-year college degree graduate
\item  4-year college degree graduate
\item  Masters or doctoral degree graduate
\end{itemize}

What do you consider to be your race? Select all that apply
\begin{itemize}
\item  American-Indian or Alaska Native
\item  Asian
\item  Black or African American
\item  Native Hawaiian or other Pacific Islander
\item  White
\item  Other
\end{itemize}

Are you of Hispanic, Latino, or Spanish origin?
\begin{itemize}
\item  Yes
\item  No
\end{itemize}

What gender do you identify with?
\begin{itemize}
\item  Female
\item  Male
\item  Transgender
\item  Other
\end{itemize}

What is your age?
\begin{itemize}
\item  21 years or younger
\item  22-35 years old
\item  36-45 years old
\item  46-55 years old
\item  56-65 years old
\item  66-75 years old
\item  76 years or older
\end{itemize}

Which of the following best describes your current employment status?
\begin{itemize}
\item  Employed for wages
\item  Self-employed
\item  Homemaker
\item  Unemployed
\item  Retired
\item  Unable to work
\end{itemize}

How much do you agree with the following statement: "Artificial intelligence can have a beneficial impact in medicine."
\begin{itemize}
\item  Strongly Agree
\item  Somewhat Agree
\item  Neither Agree nor Disagree
\item  Somewhat Disagree
\item  Strongly Disagree
\end{itemize}

\subsection*{Health Literacy Test}

Participants were given the following instruction "Before we begin the study, we want to understand how much you already know about cancer. Please answer these questions WITHOUT looking at any other resources or websites.". These questions are from the CHLT-6 cancer literacy instrument \cite{dumenci2014measurement}.

The normal range of hemoglobin for a male is 13.3-17.2 g/dl. Joe's hemoglobin is 9.7 g/dl. Is Joe within the normal range?
\begin{itemize}
\item  Yes
\item No
\end{itemize}

A biopsy of a tumor is done to ...
\begin{itemize}
\item Remove it
\item Diagnose it
\item Treat it
\end{itemize}

If a patient has stage 1 cancer, it means the cancer is ...
\begin{itemize}
\item Localized
\item In nearby organs
\item In distant sites
\end{itemize}

The role of a physical therapist is to talk to a patient about emotional needs.
\begin{itemize}
\item True
\item False
\end{itemize}

A tumor is considered "inoperable" when it cannot be treated with ...
\begin{itemize}
\item Radiation therapy
\item Surgery
\item Chemotherapy
\end{itemize}

Sally will get radiation therapy once a day, Monday through Friday. If Sally has therapy for 4 weeks, how many times will she get radiation therapy?
\begin{itemize}
\item 5
\item 15
\item 20
\end{itemize}

\subsection*{Screenshots of Survey}

\begin{figure}[h]
    \centering
    \includegraphics[width=\textwidth]{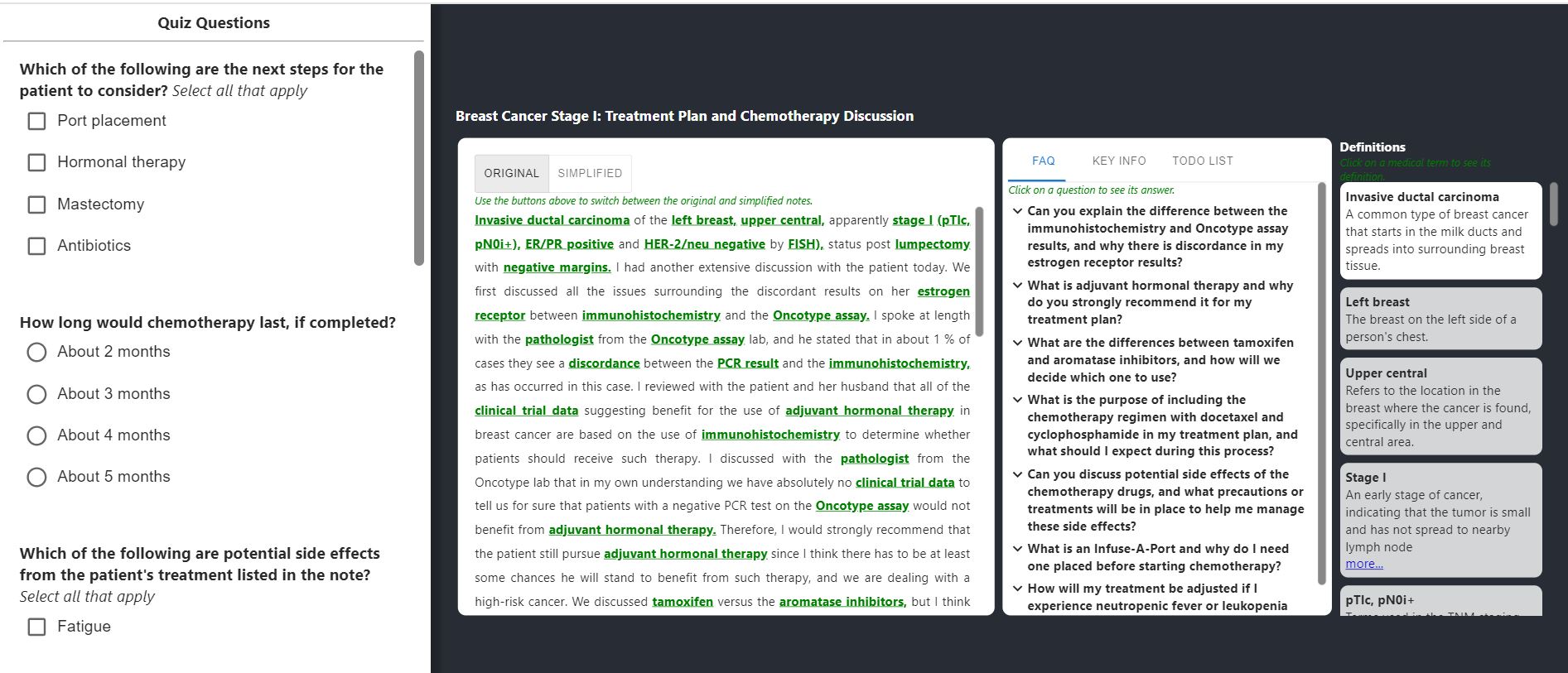}
    \caption{All augmentations condition interface with quiz questions.}
    \label{fig:sc1}
\end{figure}

\begin{figure}[h]
    \centering
    \includegraphics[width=\textwidth,height=0.9\textheight]{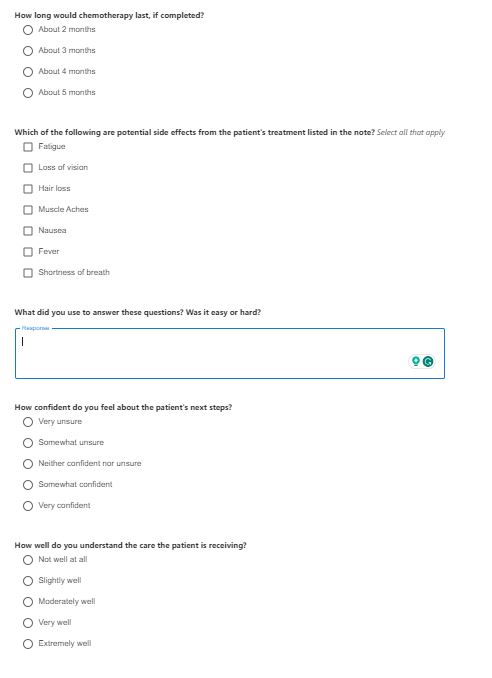}
    \caption{Quiz questions for an example note.}
    \label{fig:sc1}
\end{figure}

\begin{figure}[h]
    \centering
    \includegraphics[width=\textwidth]{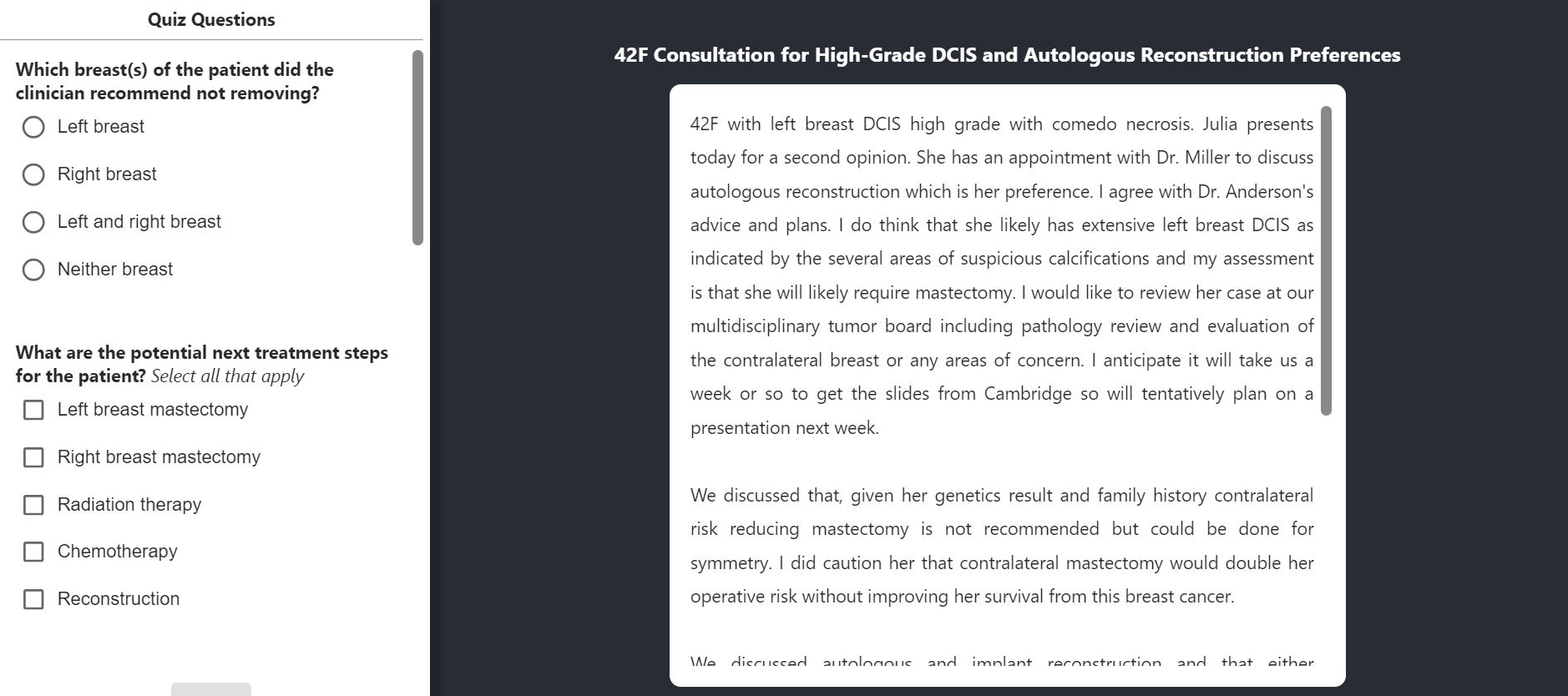}
    \caption{Control condition interface with quiz questions.}
    \label{fig:sc1}
\end{figure}

\begin{figure}[h]
    \centering
    \includegraphics[width=\textwidth]{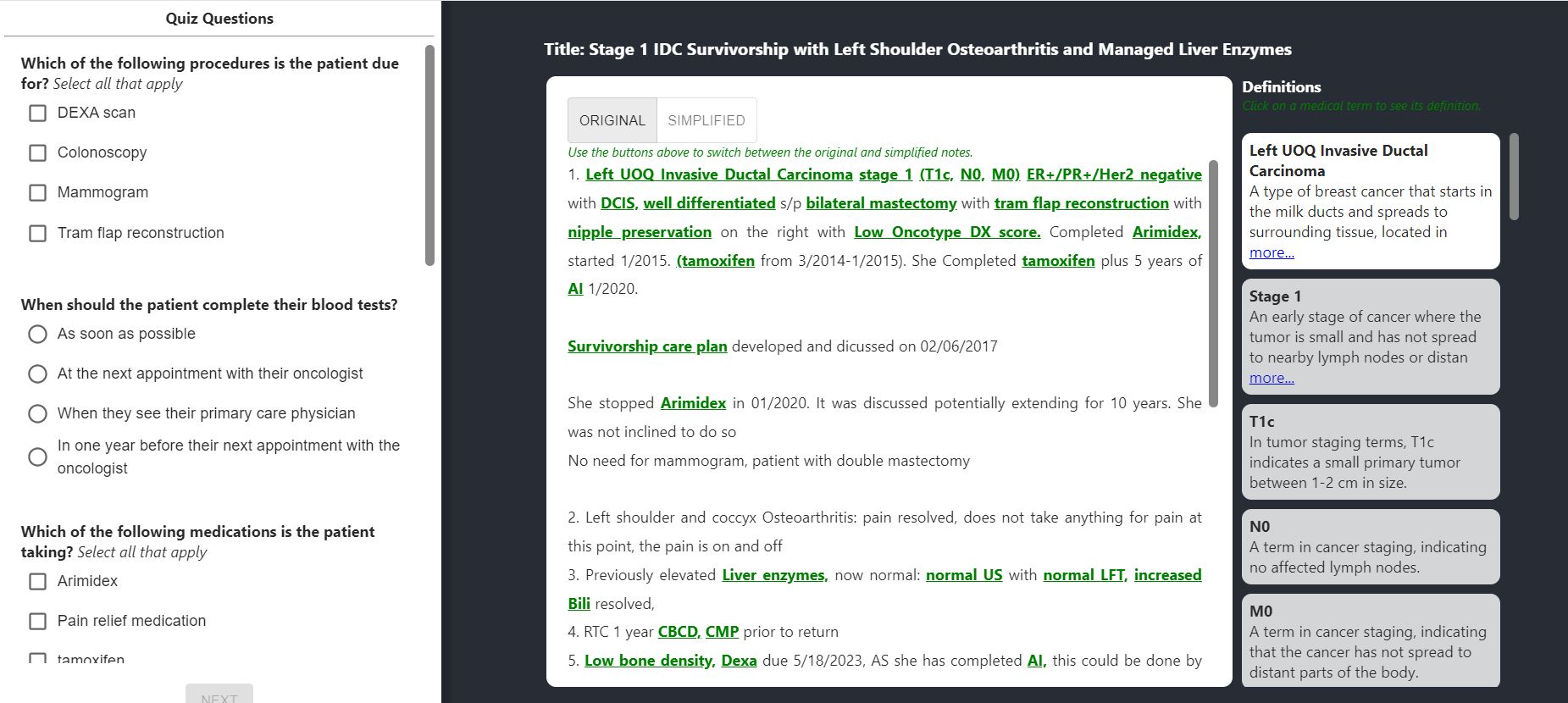}
    \caption{Select augmentations condition interface with quiz questions.}
    \label{fig:sc1}
\end{figure}

\begin{figure}[h]
    \centering
    \includegraphics[width=\textwidth]{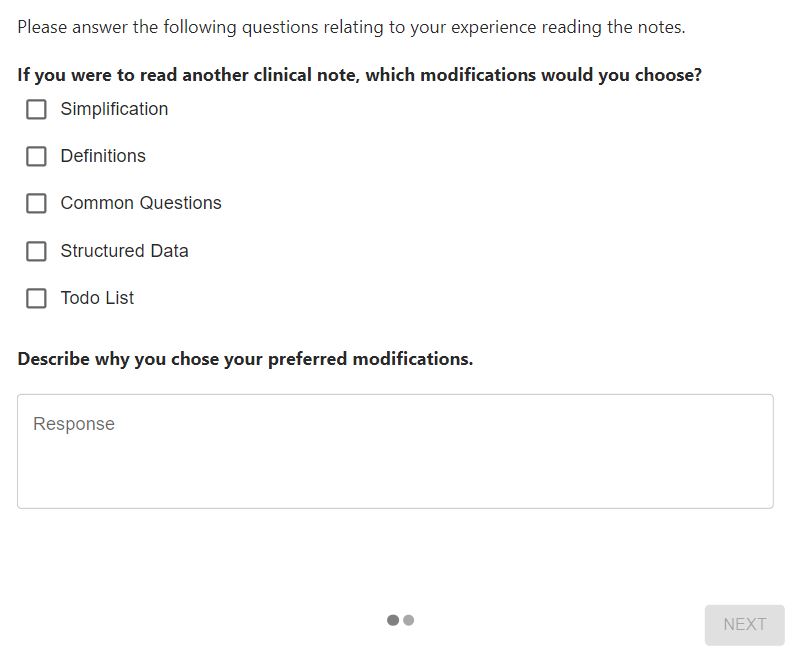}
    \caption{Post survey question.}
    \label{fig:sc1}
\end{figure}

\begin{figure}[h]
    \centering
    \includegraphics[width=\textwidth]{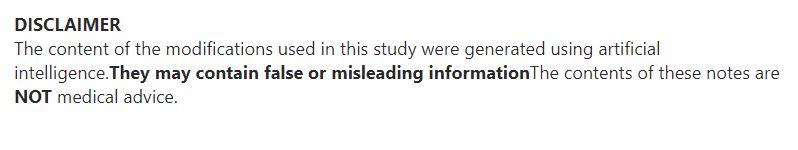}
    \caption{Disclaimer at the end of the survey.}
    \label{fig:sc1}
\end{figure}